# Design and CT imaging of Casper, an anthropomorphic breathing thorax phantom

Josie Laidlaw[1], Nicolas Earl[1], Nihal Shavdia[1], Rayna Davis[1], Sarah Mayer[1], Dmitri Karaman[3], Devon Richtsmeier[2], Pierre-Antoine Rodesch[2], Magdalena Bazalova-Carter[2]

[1] Department of Mechanical Engineering, University of Victoria, Victoria, British Columbia V8P 5C2, Canada
[2] Department of Physics and Astronomy, University of Victoria, Victoria, British Columbia V8P 5C2, Canada
[3] Axolotl Bioscience, Victoria, British Columbia V8W 2Y2, Canada

E-mail : prodesch@uvic.ca

**Abstract**

The goal of this work was to build an anthropomorphic thorax phantom capable of breathing motion with materials mimicking human tissues in x-ray imaging applications. The thorax phantom, named Casper, was composed of resin (body), foam (lungs), glow polyactic acid (bones) and natural polyactic acid (tumours placed in the lungs). X-ray attenuation properties of all materials prior to manufacturing were evaluated by means of photon-counting computed tomography (CT) imaging on a table-top system. Breathing motion was achieved by a scotch-yoke mechanism with diaphragm motion frequencies of 10 - 20 rpm and displacements of 1 to 2 cm. Casper was manufactured by means of 3D printing of moulds and ribs and assembled in a complex process. The final phantom was then scanned using a clinical CT scanner to evaluate material CT numbers and the extent of tumour motion. Casper CT numbers were close to human CT numbers for soft tissue (46 HU), ribs (125 HU), lungs (-840 HU) and tumours (-45 HU). For a 2 cm diaphragm displacement the largest tumour displacement was 0.7 cm. The five tumour volumes were accurately assessed in the static CT images with a mean absolute error of 4.3%. Tumour sizes were either underestimated for smaller tumours or overestimated for larger tumours in dynamic CT images due to motion blurring with a mean absolute difference from true volumes of 10.3%.

## 1. Introduction

Anthropomorphic phantoms are a key tool to enhance the development of imaging and therapeutic medical devices. They allow testing of radiological and radiotherapeutic protocols without the constraints of animal or human experiments. The rise of 3D printing in the last decade [1] has considerably facilitated and increased the sophistication of realistic anthropomorphic phantoms. To provide relevant information for human application, phantoms must mimic anthropomorphic features in terms of material properties, geometrical resemblance (dimension and details), and motion. The latter is more specifically of importance for thoracic imaging and radiotherapy procedures. Depending on the medical application, either the respiratory or the cardiac motion needs to e reproduced. The current work focuses on the development of an x-ray imaging phantom replicating the lung respiratory motion.

Dynamic (or 4D) anthropomorphic phantoms are commercially available, such as the QUASAR™ Respiratory Motion Phantom (Modus Medical Devices, London, ON, Canada) or the Dynamic Thorax Phantom (Model 008A, CIRS Incorporated, Norfolk, VA, USA). Both phantoms are made of solid parts. A controller moves a tumour or target in one or several directions to reproduce the motion amplitude of a tumour or a calcification. The lung motion frequency is well simulated, but the motion and body features remain simple: round body shapes and spherical tumours. 3D printing can enable the manufacturing of solid parts with finer details, and it was mainly applied to develop static lung phantoms with novel features. For example, it can produce a geometry with biological variability, e.g with heterogeneous material density [1], a geometry based on standard simulation phantoms like the XCAT phantom [2] or based on patient data. In this last case and depending on the original patient choice, different configurations can be reproduced like a healthy patient [3] or a postmastectomy female patient [4]. Latest 3D printing techniques can reproduce patient anatomy down to 0.1 mm details, enabling the reproduction of a lung infected by COVID-19 pneumonia [5]. 3D printing was also used to manufacture dynamic phantoms with more complex shapes than commercially available 4D phantoms, e.g, with non-spheric tumours [6] or with a fully anthropomorphic thoracic cage [7], [8]. However, most 3D printers can only manufacture solid parts resulting in non-deformable lungs.

The optimal way to reproduce human lung motion is to allow for lung deformation. Two different possibilities have been investigated in the literature to manufacture deformable human lungs: airtight bags and foam. Foam has the advantage to better reproduce human lung attenuation [9] whereas air underestimates it. Moreover, the density of the foam can be adapted to reproduce either healthy or pathological lungs [9]. With some modifications, a contrast agent can also be injected in the foam [10] to match a specific radiological protocol. Whereas an airtight bag will simply fill the space in a phantom, the foam must have a lung shape to fit into an anthropomorphic thorax structure, leading to a more complicated manufacturing process. As a result, most of the published works using foam deformable lungs do not have a full thoracic cage but only a cylindrical shape compressed by a pushing plate [10]–[12]. In contrast, airtight bags have been used to reproduce lung inflation. They are made of a flexible and low attenuation material (e.g silicon) and air is pumped in it [13]–[15], or alternatively water is pumped around it [16].

Another possibility to ensure a tumour motion in a 3D printed thoracic cage is a rod translating into a deformable material as sawdust [17] or foam [18]. In the human body, the breathing motion is primarily achieved with the diaphragm [19], [20]. The diaphragm is a torso membrane located below the lungs and its movement (mainly a superior-inferior (SI) translation) triggers the thoracic dilation and contraction. A dynamic phantom with the motion ensured by a torso membrane was developed by Ranjbar *et al.*, 2019 [21] with a simplified anatomy: the thoracic cage was filled with a foam which was not lung shaped and the torso membrane was a plate. Bolwin *et al.* [22] developed a phantom with anthropomorphic lungs and diaphragm but without a solid bone structure (no ribs or spine). A brief summary of developed anthropomorphic phantoms is listed in **Table 1**. In our work, a dynamic phantom with a 3D printed bone structure (ribs and spine) was developed. The lung motion is ensured by the translation of a torso membrane (shaped as a realistic diaphragm) compressing two lung shaped foam parts. This enables a realistic replication of the lung motion, in an anthropomorphic thorax phantom, while using materials replicating the human body x-ray attenuation.

**Table 1:** Bibliographical table of recent published phantom reproducing the respiratory contrast and motion.

| Author | Anthropomorphic feature | Deformable lung Yes/No (material) | Motion | Use of 3D printing Yes/No (Item 3D printed) |
|---|---|---|---|---|
| Kostiukhina 2017 [6] | Thoracic cage (ribs and soft-tissue) | No (high density balsa wood) | Superior-Inferior translation and rotation of a cylinder rod | Yes (complex tumour shape) |
| Pallotta 2018 and 2019 [7-8] | Thoracic cage (ribs and soft-tissue) | No (cork) | 3D translation of the cork blocks | Yes (3D soft-tissue cage) |
| Lehnert 2019 [11] | One lung | Yes (lung-shaped foam) | No | No |
| Gholampourkashi 2020 [12] | One lung | Yes (cylindrical foam) | A diaphragm compressing the foam | No |
| Shin 2020 [10] | One lung | Yes (cylindrical foam) | A diaphragm compressing the foam | Yes (flexible airways) |
| Polycarpou 2015 [13] | Thoracic cage (ribs and soft-tissue) | Yes (silicone bags) | Air pumped in lungs | No |
| Perrin 2017 [14] | Thoracic cage (ribs and soft-tissue) | Yes (polyurethane bags) | Air pumped in lungs | No |
| Matsumoto 2019 [51] | Thoracic cage (ribs and soft-tissue) | Yes (urethane bags) | Air pumped in bags | No |
| Black 2021 [16] | Thoracic cage (ribs and soft-tissue) | Yes (silicone bags) | Water pressured in thorax | No |
| Mayer 2015 [17] | Thoracic cage (ribs and soft-tissue) | Yes (sawdust bag) | 3D translation of a tumour | Yes (ribcage and soft tissue) |
| Colvill 2020 [18] | Thoracic cage (ribs and soft-tissue), liver (silicon), arteries | Yes (3D foam printed lung) | Superior-Inferior translation of a cylinder rod deforming the 3D foam printed lung | Yes (3D foam printed lung, vein-like structures within the liver) |
| Ranjbar 2019 [21] | Thoracic cage (ribs and soft-tissue) | Yes (not lung shaped foam) | Torso membrane translation | No |
| Bolwin 2018 [22] | Thoracic cage (only soft-tissue), heart left ventricle, liver | Yes (silicone) | Torso membrane translation | No |
| **Casper** | Thoracic cage (ribs and soft-tissue) | Yes (Lung shaped foam) | Torso membrane translation | Yes (ribs, spine and thorax shaped mould) |

## 2. Materials and Methods

2.1 Overall aspect

Casper, the phantom developed in our work, consisted of a simplified anthropomorphic thorax with four main components: ribs, lungs, lung tumours, and surrounding soft tissue (cf **Table 1**). The materials selected for these components were chosen to have similar x-ray properties to their human counterparts, which was verified by x-ray computed tomography (CT) imaging using a table-top photon-counting CT scanner (PCCT). The phantom incorporated lung breathing motion from a diaphragm-mimicking scotch-yoke mechanism. The scotch-yoke mechanism moved to expand and compress the lungs, simulating real breathing. Various sizes of lung tumours were placed in the lungs to examine the ability of an x-ray imaging modality to clearly image the tumours while the lungs were moving.

With permission, the phantom was designed in Python based on the existing XCAT phantom [23], which was created for imaging research to improve imaging techniques and equipment. An online CT scan file converter Embodi3D was used to convert the modified XCAT phantom into STL files (a file format native to the stereolithography computer-aided design (CAD) software), which could then be used in CAD and 3D printing programs.

2.2 Material samples x-ray attenuation properties

X-ray attenuation properties of phantom materials replicating the various tissues were evaluated using a bench-top photon-counting CT (PCCT) system. The system consisted of an x-ray tube (MXR 160/22, Comet Technologies, San Jose, CA), a photon-counting detector (PCD) (Redlen Technologies, Saanichton, BC), and a rotation stage (Newport Corporation, Irving, CA) positioned between the PCD and x-ray tube. All components were mounted on linear motion stages allowing for vertical or horizontal translation, or both. The PCD crystal was 2 mm thick cadmium zinc telluride (CZT) with an active area of 8 x 190 mm and a pixel pitch of 330 μm.

All CT acquisitions of the tissue materials were performed at a tube voltage of 120 kVp, a tube current of 2 mA, using the small focal spot of 1 mm, and with 5 mm aluminium filtration mimicking clinical scanners. The source to detector distance was 578 mm and the source to isocentre distance was 322 mm. 720 projections were acquired over 360° while the rotation stage was rotated at 2° per second. The six PCD energy thresholds were set to 35, 45, 55, 70, 90, and 120 keV and images were reconstructed using all energies from 35-120 keV. Detailed projection correction algorithms are presented in our previous work [24], [25].

CT images were reconstructed using the Feldkamp-Davis-Kress (FDK) algorithm with a Hamming filter [26], which was implemented in the TIGRE python package [31]. The full data set consisted of 24 slices of 512 x 512 pixels, with an image size of 105 x 105 mm. The *z*-coverage of the 24 slices was 4.41 mm. Once the CT images were reconstructed, they were normalized to CT numbers expressed in Hounsfield units (HU) using the equation:

$$HU = 1000 * \left(\frac{\mu - \mu_{water}}{\mu_{water}}\right)$$

where μ is the signal in each individual pixel of the reconstructed image, and $\mu_{water}$ is the signal of water, taken as the average signal in a water vial scanned in the same image.

For each of the materials evaluated for x-ray attenuation properties, the samples were placed on the rotation stage along with a vial containing water (cf **Figure 1**). The HU signal was evaluated in the reconstructed CT images as the mean of a region-of-interest (ROI) taken from within the sample. The HU value given by this mean signal was then compared to the HU value of the actual tissue type that is seen in clinical CT images.

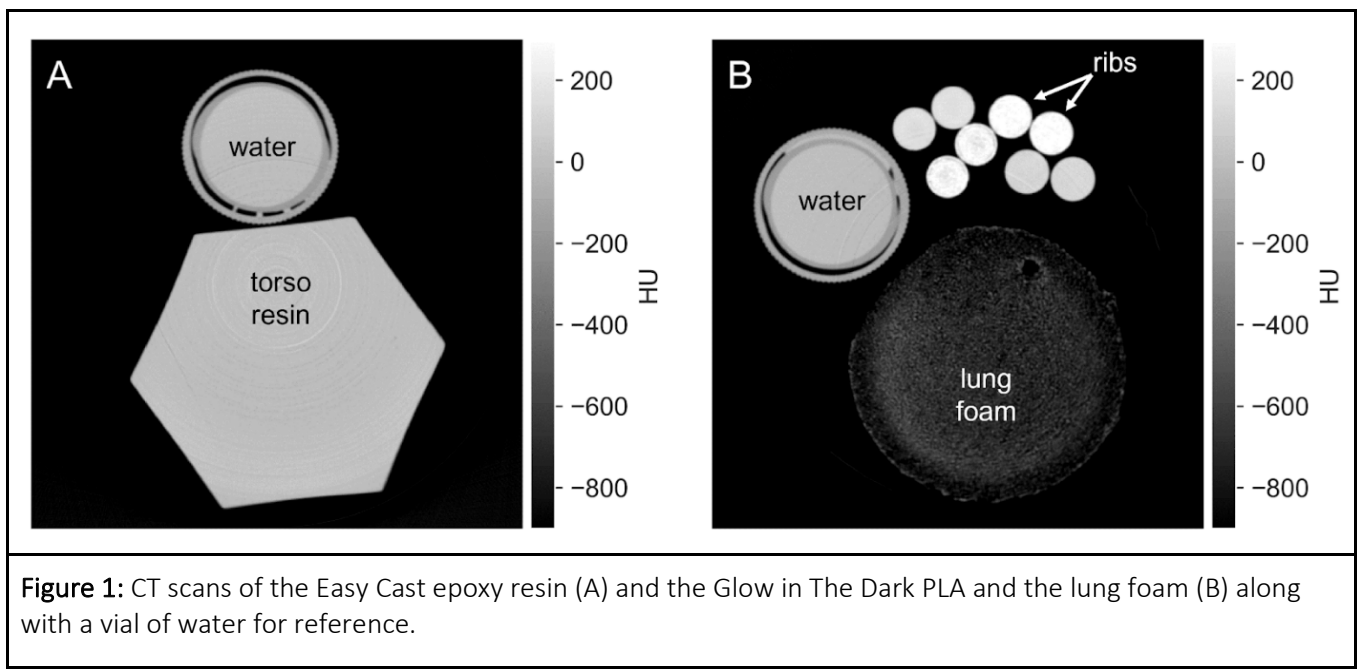

**Figure 1:** CT scans of the Easy Cast epoxy resin (A) and the Glow in The Dark PLA and the lung foam (B) along with a vial of water for reference.

2.3 Phantom materials

Materials were selected that would replicate human soft tissue, bone, lung tissue, and tumours when scanned with x-ray CT. Research was conducted to find and image materials with suitable x-ray attenuation.

*Torso:* Casper's torso was designed to mimic soft tissue surrounding the ribs and lungs. A study by Alshipli *et al*. [27] found epoxy resin to have low CT numbers averaging at 69.2 HU ± 6.1. As this value was close to the target of 0 HU, four samples of different epoxy resins were scanned and compared to a sample of water. Of the four scanned samples, Easy Cast epoxy resin (Environmental Technology Inc., Kalamazoo, MI, USA) was found to have the most desirable radiodensity with a result of 5 HU ± 2 HU and was selected as the soft tissue material (**Figure 1A**).

*Ribs:* Materials for the ribcage were limited to 3D-printable filaments. The XCAT data [23] provided attenuation coefficients for the ribs and spine. Target CT numbers were therefore determined to be 358 HU for the spine and 788 HU for the ribs. Only a few 3D-printable materials were found to have high HU values. Ceh *et al*. [28] reported that glow polyactic acid (PLA) was found to have significantly higher CT numbers than other accessible filaments. Three glow PLA samples were tested, with HUs ranging from 275-325 HU. The selected material, Glow in The Dark PLA Filament (Duramic 3D, Shanghai, China), printed at 110% infill had CT numbers of 292 HU (**Figure 1B**).

*Lungs:* A lung material was required that had similar CT numbers to human lungs and could be easily compressed to simulate breathing motion. The target CT number for lungs from the XCAT data was calculated to be -700 HU. Standard polyurethane foam was selected as it could be poured into a mould while in a liquid state, then expanded into a solid foam to achieve the desired shape. The selected foam was found to be –840 HU at rest and –680 HU when compressed (**Figure 1B**).

*Tumours:* Xie *et al.* [29] found that cancerous tumours in the lung region had very similar CT numbers to natural PLA, where tumours have CT numbers of approximately 111 HU, and the PLA was 128 HU. Matter3D Performance PLA (Matter3D, Victoria, BC, Canada) was therefore selected and was determined to have CT numbers of 87 HU.

2.4 Manufacturing processes

The primary manufacturing processes used in this project were moulding and additive manufacturing.

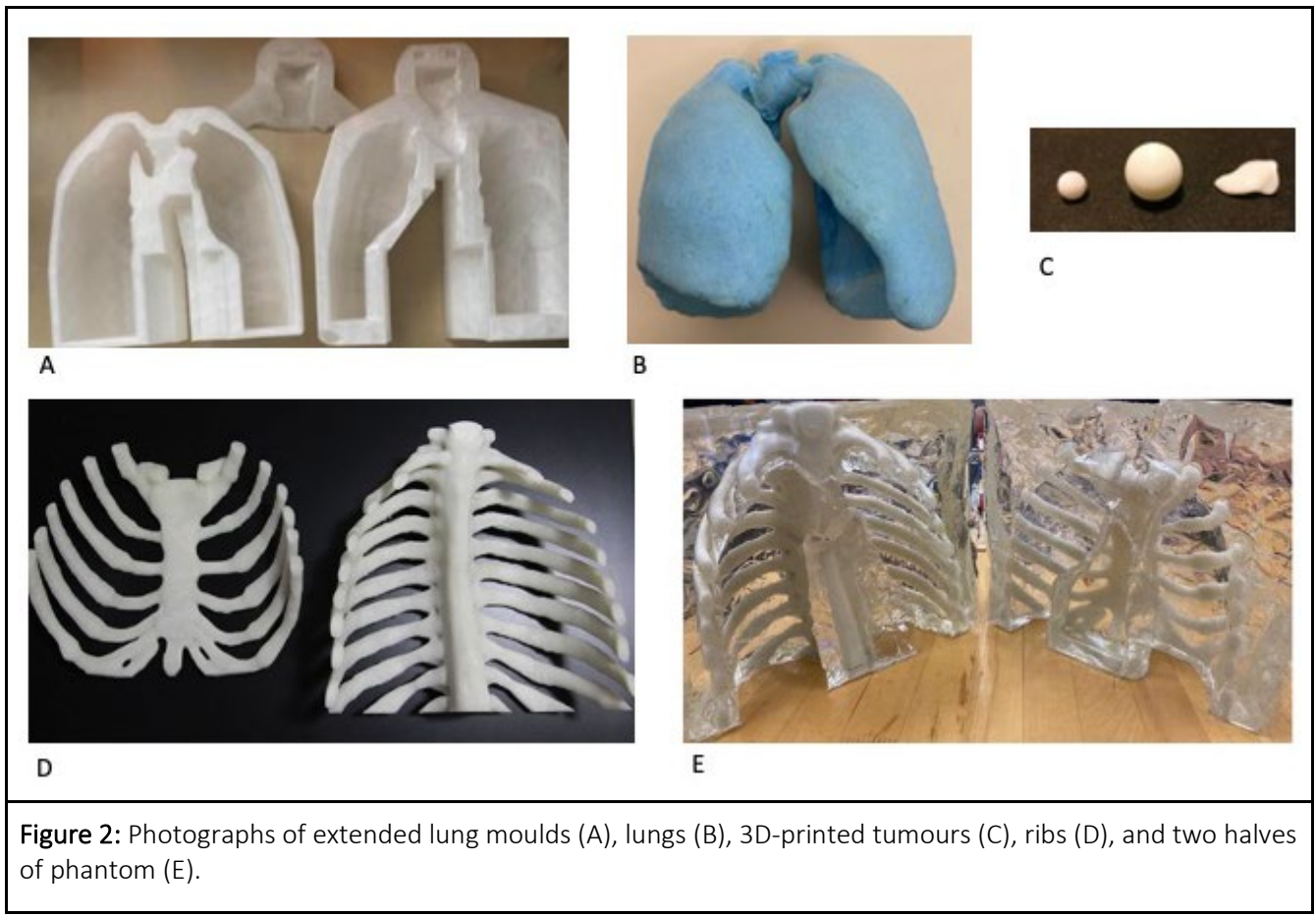

**Figure 2:** Photographs of extended lung moulds (A), lungs (B), 3D-printed tumours (C), ribs (D), and two halves of phantom (E).

Moulds were used to create Casper's soft tissue, lungs and lung-cavity place holders; while 3D-printing was used to manufacture the ribcage, tumours and moulds.

*Lungs:* PLA moulds were created and Turtle Wax (Turtle Wax Inc, Addison, Illinois, USA) was used as a release agent, as it was accessible and the process repeatable. These moulds were designed in two halves split at the plane of the centre of the trachea, as well as perpendicularly on one of the sides, to allow for easy removal of the lungs (**Figure 2A**). When forming the lungs, the polyurethane foam mixture was poured through two openings made at the top of the moulds to allow it to expand and fill the entire mould when clamped together. After expansion, the excess foam was trimmed off. The final lungs are shown in **Figure 2B**.

*Tumours:* Tumours of various sizes were 3D printed with Matter3D Performance PLA, including spheres with the following diameters: 0.5, 1.0, 1.5 and 2.0 cm, and an irregular shaped tumour (around 2.0 cm long) from real tumour CT data (**Figure 2C**). The tumours were printed at 100% infill, and fine threads were attached to the tumours mid-print to suspend the tumours at different locations during the lung manufacturing.

*Ribs:* The 3D-model of a ribcage was obtained from existing XCAT data and was initially separated into the front and rear section of the ribcage with the software Cura (Ultimaker, Utrecht, the Netherlands). This initial separation was due to the manufacturing of the torso, where the two halves of the body were assembled. These two halves were further separated due to limitations in the printer volume. Plane cuts were used as they were found to assemble with fewer inner voids than other assembly methods. Small detached rib components were printed separately.

For each of the rib cage sections printed with Glow in the dark PLA Filament, a 0.6 mm nozzle was used with 0.4 mm layer height, 110% concentric infill, 105% extrusion multiplier, and 75% of the standard print speed. The largely vertical and narrow ribs, mixed with the higher extrusion multiplier required lower speeds to print without issues. Support structures were kept at a default and provided sufficient support to overhangs. After all of the prints were completed, the connecting surfaces were sanded down until flat and glued using a UV-cured resin. With the final assembled ribcage, sanding was done with a Dremel to create a smoother surface and remove leftover support material. Finally, a heat gun was used to gently bend any warped ribs back into place (**Figure 2D**).

*Surrounding Chest:* The manufacturing process of the chest involved designing a process that would create the geometry that replicates the XCAT data while encasing the ribs in the soft tissue and leaving cavities for the lungs. The lungs required cavities to ensure there was sufficient room to be compressed by the diaphragm to accurately simulate breathing.

The chest mould was printed in four parts due to size constraints of 3D printers. Two parts were connected using dowel alignment features to create a front and back half. The moulds were coated in release agents as well as packing tape and plastic wrap to ensure the resin separated from the mould.

To create the cavity for the polyurethane foam lungs, an extended set of lung moulds and extended lungs were designed. The extended lung moulds continued down from the bottom of the original lung mould, creating a longer shape that was flat on the bottom. This created room for the connection mechanism to compress the bottom of the lungs.

The resulting extended lung halves were attached to a sheet of etched plexiglass, along with the ribs, which were then placed face down into the soft tissue mould to act as a lid for the resin pour. The plexiglass was etched with the pattern of the rib cuts and location of the lungs using a laser cutter. Once the resin cured, the plexiglass was removed, leaving the ribcage encased in the resin. The firm placeholder lungs were then removed, revealing the extended lung cavity, with room for the original lungs and the connection mechanism to be inserted.

The resin was poured in small quantities to avoid overheating and risking damage to the ribs or the moulds. Upon removal of the moulds, the phantom consisted of the front and back halves, which could then be connected to form the full phantom. **Figure 2E** shows the two halves of the phantom after their removal from the body moulds. The phantom halves then required post-processing. Packing tape was removed, and the inside faces were faced on a mill and sanded for a smooth finish. The outside was then given a thin layer of resin to fill any significant imperfections. All inside and outside edges were sanded, and the phantom was polished with oil.

2.5 Motion controller

The breathing mechanism was created using a small circuit with an Arduino Mega 2560 (Arduino, Victoria, BC) that was used to power a 12-volt DC-motor (McMasterCarr, Los Angeles, CA). The circular motion from the motor was translated to SI motion using a shaft and a scotch-yoke mechanism. This sinusoidal motion then was applied to the lungs using a connection mechanism to replicate the diaphragm movement to achieve a natural breathing rate. The connection mechanism with two pieces of resin pushed up on the underneath contours of the lungs. Three rotation speeds of 10, 15, and 20 rpm and three translation displacements of 1, 2, and 3 cm could be set.

The motion controller was installed in a wooden box placed inferior to the thorax phantom, outside of the field of view of interest (cf **Figure 3**). An additional light support was designed to sustain the phantom in a lying position (cf **Figure 4B**).

2.6 X-ray imaging

The final phantom was scanned in a clinical CT scanner in the horizontal (lying) position and in a research facility radiography imaging system in the vertical (standing) position. When moving, the diaphragm translation extent was set to 2 cm. The diaphragm in its minimal SI z-position was set as the z-origin.

*Radiography:* Casper was set in a vertical position for radiography imaging. X-ray imaging was performed with a CsI scintillator flat panel detector (XRpad2 3025HWC) with a 0.1 mm pixel pitch. The source was placed at 876 mm from the detector and 622 mm from the phantom. Two measurements were made at the minimum and maximum z-position in order to evaluate the offset in the vertical position. The minimum and maximum z-position are representing, respectfully, the inspiration and expiration peaks.

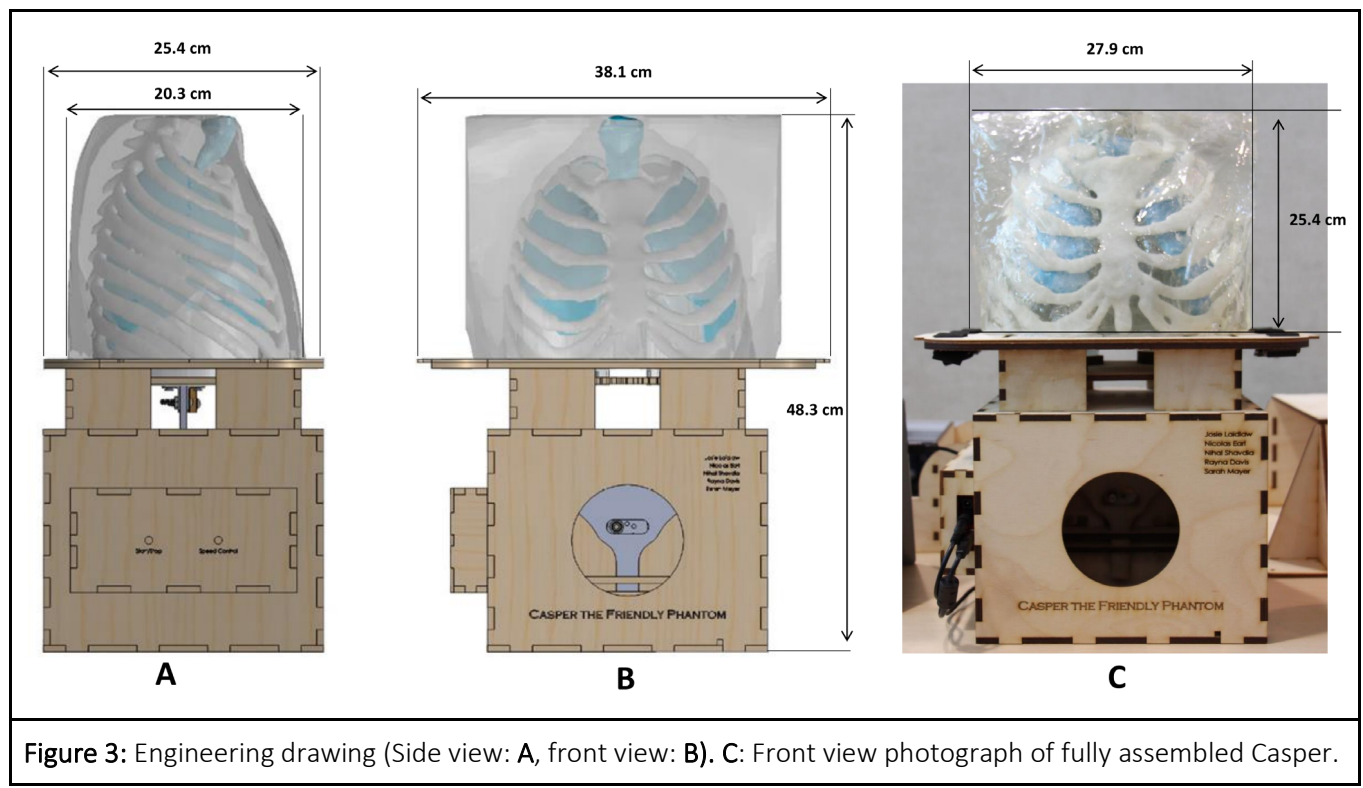

**Figure 3:** Engineering drawing (Side view: **A**, front view: **B**). **C**: Front view photograph of fully assembled Casper.

For both measurements ROIs were manually placed for each tumour and for the torso membrane. The centroid SI position difference was evaluated as a pixel number. Because of the magnification and radiographic mode, the offset measurement needed to be calibrated. As the torso membrane is rigid and directly attached to the motor, its displacement was exactly 2 cm. This distance was used to convert the pixel numbers into equivalent length.

*CT imaging:* The phantom was imaged with 120 kVp and 10 mAs with a 16-slice Optima 580 CT scanner (GE Healthcare, Chicago, IL). It was laid on the CT system table and scanned statically mimicking a breath hold CT, at the inspiration and expiration peaks to estimate each tumour displacement. It was also scanned dynamically to replicate the impact of the breathing movement on the gross tumour volume (GTV) estimation [30]. All CT images were reconstructed with a 'body' filter and a 500 mm field-of-view, a 512 matrix and a 2.5 mm slice thickness.

In the three reconstructed images, tumours were delineated with a threshold at -350 HU. For each tumour were measured the volume and z-centroid. The latter was used to evaluate the tumour displacement defined as the z-centroid difference between the minimum and maximum z-position. ROIs were also placed in the surrounding resin (soft tissue), rib, vertebrae, tumour, and lung in order to measure final CT numbers.

## 3. Results

### 3.1 Assembled phantom

The fully assembled Casper is shown in **Figure 3**. Its dimensions (20.3 x 27.9 x 25.4 cm) correspond well to a small adult thorax. To facilitate transportation and enables modularity, Casper is easy to assemble/dissemble thanks to its two halves design.

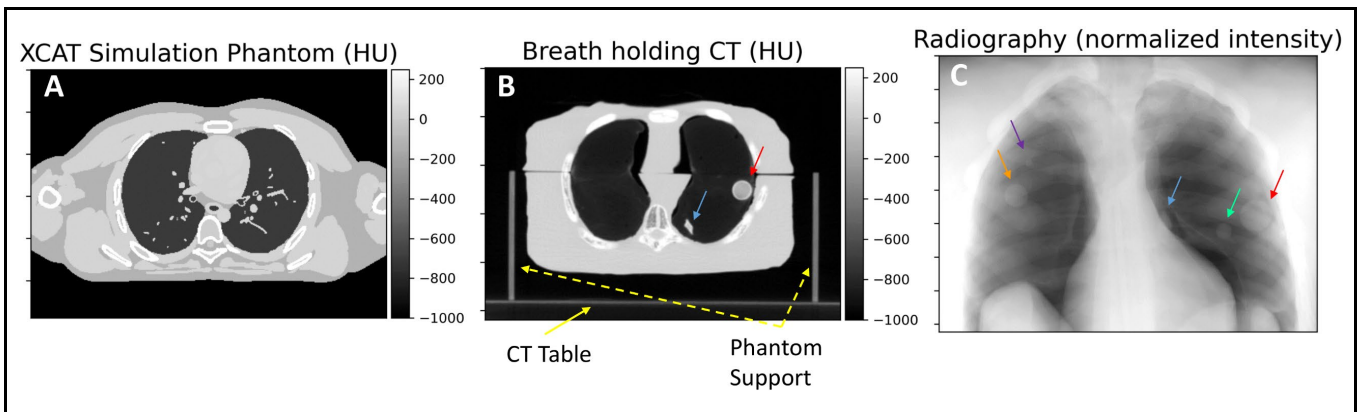

**Figure 4:** XCAT numerical phantom (A), Casper static CT image (B) and Casper vertical position radiography (C). The arrows are indicating the tumours' locations. Purple, green, orange and red arrows point at 0.5, 1.0, 1.5 and 2.0 cm diameter tumours, respectively, and the blue arrow points to the tumour with a complex shape.

## 3.2 Static CT images

**Figure 4** displays the static CT and radiography images. The tumours are located at various SI and in-plane locations. The phantom reproduces well the XCAT overall shape and the global attenuation range of a human chest. However, the attenuation is locally underestimated for the tumours, as the HU numbers in **Table 2** show.

**Table 2:** HU numbers measured in the CT stationary image compared to a healthy patient.

| Body Area | Casper | Average Patient |
|---|---|---|
| Soft tissue | + 46 | - 30 / 40 |
| Tumour | - 125 | 20 / 70 |
| Lung | - 840 | - 800 |
| Rib | + 140 | + 125 |

## 3.3 Dynamic CT images

A coronal view of the CT images acquired with the motion deactivated and activated are presented in **Figure 5**. Motion artefacts are visible in the connection mechanism and in the blur of the lungs. The 2-cm diameter tumour shape is distorted in the motion direction and no longer appears as a sphere.

The volumes of each tumour measured statically and dynamically are summarised and compared to the theoretical volumes in **Table 3**. Given the low outer plane resolution, the volume estimation is highly sensitive to the partial volume effect and the selected threshold to delineate tumours. The impact of the motion artefacts is clearly visible with a volume underestimation of the smaller tumours and an overestimation of larger tumours. The mean absolute error of volume estimation was 4.3% and 10.3% for the static and dynamic CT images, respectively.

**Table 3**: Theoretical and measured volume for static and breathing Casper. N/A: Non Available: the complex shaped tumour was not defined as a sphere but has a theoretical volume between a 1.0 and 1.5 cm diameter sphere.

| Tumour diameter | Tumour volume (mm$^3$) | | |
|---|---|---|---|
| | Theoretical | Static | Breathing |
| 0.5 cm | 65 | 67 | 64 |
| 1.0 cm | 524 | 527 | 441 |
| N/A | 1113 | 997 | 1092 |
| 1.5 cm | 1767 | 1640 | 1545 |
| 2.0 cm | 4189 | 4189 | 5012 |

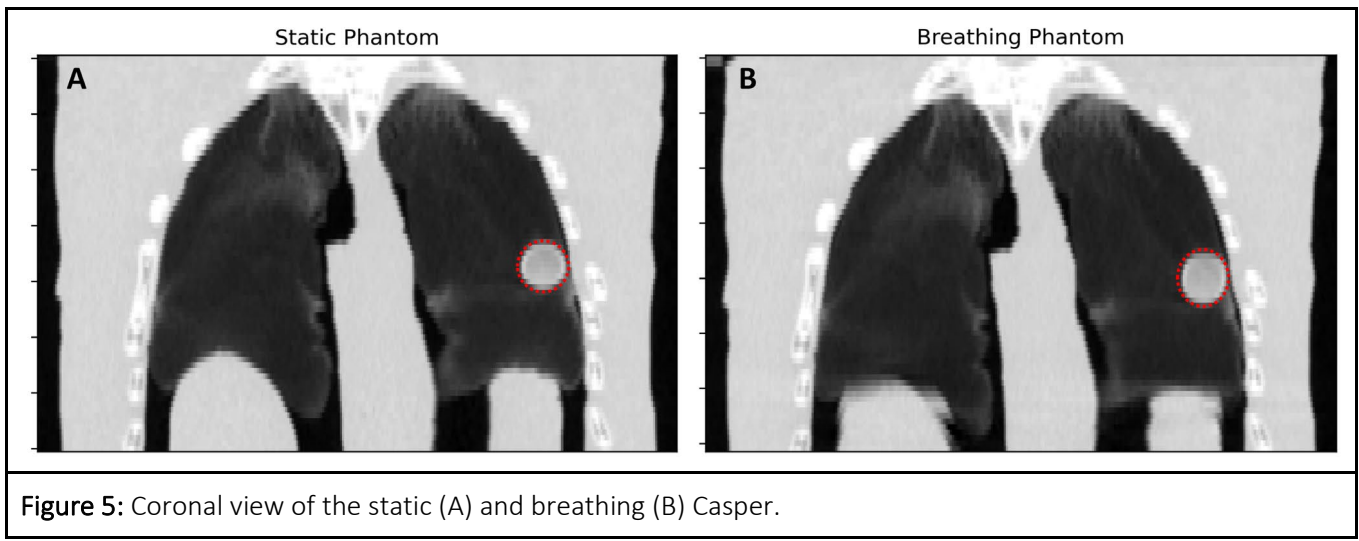

**Figure 5:** Coronal view of the static (A) and breathing (B) Casper.

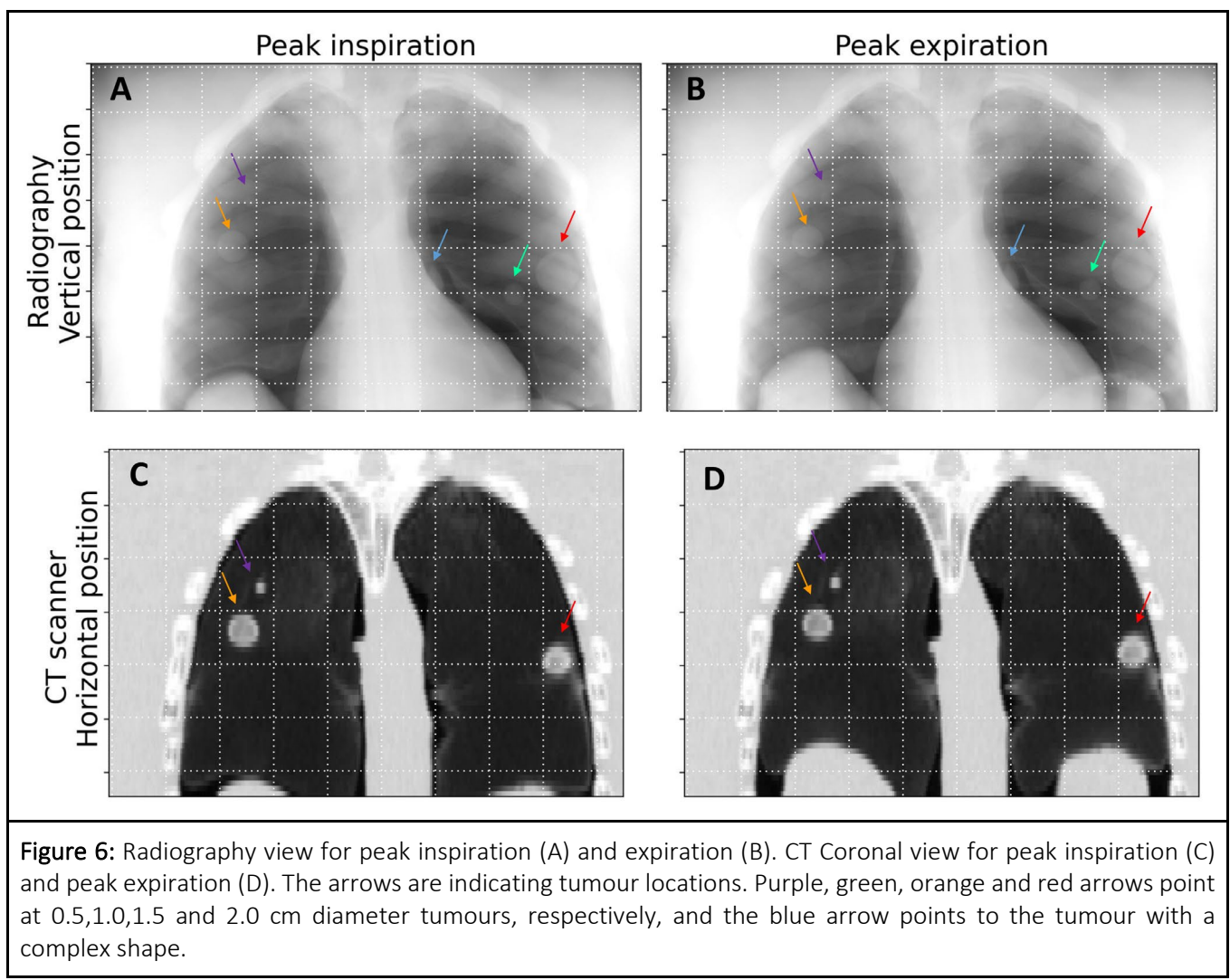

**Figure 6:** Radiography view for peak inspiration (A) and expiration (B). CT Coronal view for peak inspiration (C) and peak expiration (D). The arrows are indicating tumour locations. Purple, green, orange and red arrows point at 0.5, 1.0, 1.5 and 2.0 cm diameter tumours, respectively, and the blue arrow points to the tumour with a complex shape.

### 3.4 Tumour maximum displacement

Static radiography images and coronal CT views of Casper in its minimum and maximum diaphragm displacement are displayed in **Figure 6** and tumour offsets are presented in **Figure 7**. The offset was in general smaller in the radiography images acquired with Casper vertical than in CT images acquired with Casper horizontal. The offset was different for each tumour and linearly decreased with increasing the minimum z-position. The tumour with a complex shape was located close to the spine (**Figure 4**) and resulted in a smaller offset (0.2/0.3 cm).

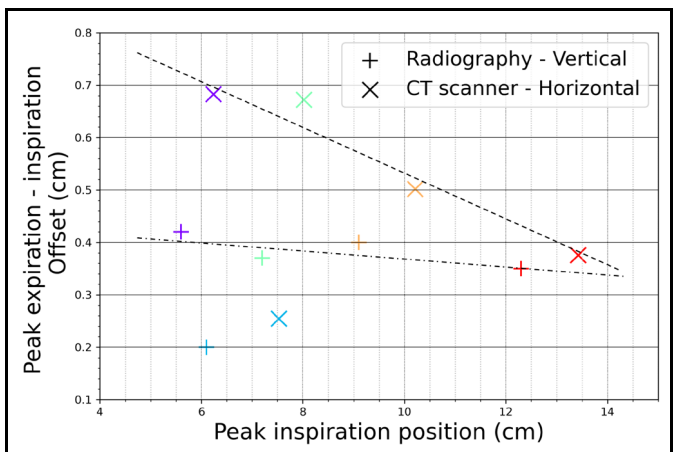

**Figure 7:** Tumour SI offsets between the inspiration and expiration peaks as a function of its SI position at the inspiration peak. The diaphragm served as the reference and was located at $z = 0$ cm. The purple, green, orange and red markers are the 0.5, 1.0, 1.5 and 2.0 cm diameter tumours, respectively, and the blue data point is the tumour with a complex shape.

## 4. Discussion

### 4.1 Manufacturing processes

Casper was successful at emulating tumour motion in lungs, however there were a few aspects of the phantom that could be improved upon. The resin for the body reached a high temperature during curing, which resulted in distortion of the bottom two ribs due to the heat. The resin also had a very long curing time between layers (8 hours), and because the resin needed to be poured in thin 1 cm-thick layers due to the high curing heat, it took 3 days to make the resin body, with pours occurring every 8 hours.

The PLA that was used to 3D print the ribs did not result in a consistent cross-sectional image of the ribs when scanned (**Figure 5**), despite using 110% infill settings on the 3D printer. In addition to this, the high-infill and extrusion printing settings, as well as the abrasive additives in the filament, led to many failed prints that were the result of warped prints and damage on the extruder gear of the printer.

A part of the manufacturing process that executed poorly was the moulding process of the resin. One reason for this is because of the use of brittle 3D printed female moulds, which had an undesirable and rough finish, increasing the possibility of bonding of the resin to the plastic. Consequently, significant amounts of tape and plastic wrap were used to prevent the resin from bonding to the PLA. The resin leaked through the tape and plastic wrap layer in some places, and the plastic wrap also resulted in a surface that did not appear smooth, and much of the geometry from the mould was lost. A significant amount of time was spent removing the phantom body from the moulds it had partially bonded to and touching it up as a result.

An additional problem involved the inconsistency of the tumour placement. The PLA tumours were placed in the lung mould by applying tension to threads that were connected to the PLA tumours and securing the threads to the outer surface of the mould. However, the rapid volume increase of the foam in the mould resulted in the strings becoming loose and the tumours getting unpredictably displaced from their original locations. This step required several tries to obtain a satisfying tumour distribution.

Additionally, the motion phase of the lungs cannot be externally measured when the breathing motion is activated. Indeed, the outside surface of the chest remained stationary as the lungs compressed internally. As a result, using an external chest marker to detect lung displacements as done for patients could not be implemented. In order to improve the performance and research capabilities of the phantom, motion tracking is desired. This could be investigated and implemented by adding potentiometers to the motor, which would allow the tracking of the motor stroke, and thus the resultant tracking of the lung compression. Alternatively, other methods of tracking such as the implementation of infrared light sources at the motor crank could be used in conjunction with an infrared camera to detect and track the motor stroke.

The manufacturing process could also be improved by selecting a resin with a lower curing temperature and a faster curing time. Although the selected EasyCast Epoxy Resin had performed well in material testing, the combination of the 3D printed rib cage and the high temperature cure resulted in some deformation of the ribs at connection points. Selecting a lower curing temperature resin that consequently cures faster would prevent the deformation of the ribs and would allow for a faster overall mould pouring process. Additionally, more testing on the ribcage material and its 3D printing properties could have been done to thoroughly investigate different infill combinations which would result in a consistent cross-sectional image of the ribs.

Another essential refinement would be the improvement of the moulds, specifically by the use of silicon moulds instead of the previously used 3D printed female moulds for the resin soft tissue pour. These can be made by 3D printing a male mould such that the silicon moulding material may be painted onto the 3D printed mould, thus creating a female silicon mould. Using a silicon mould for resin soft tissue will result in more accurate, cleaner, and reproducible results as compared to the 3D printed mould that was lined with plastic wrap and tape.

Inconsistent attenuation of the foam lungs seen if **Figures 4** and **5** may have been caused by ridges formed in the lungs due to potential gaps between the mould faces, resulting in the seepage of foam. This could be prevented by using silicon moulds with flanges that would allow for improved contact between the mould surfaces. Along with better perpendicular clamping, the flanges would prevent any seepage of the foam, thus preventing the formation of ridges on the exterior surfaces of the foam lungs.

4.2 X-ray imaging

The final Casper design accurately reproduced an adult male chest in terms of dimension and x-ray attenuation. Due to its design in two halves, it is highly modular, and lungs can be easily replaced, or further features included. An adipose-like outer shell could also be easily designed and added around Casper to simulate larger patient sizes. The wood thin phantom support and motion controller can be conveniently placed outside of the field of view in either a lying or standing position so that they do not interfere with image data acquisition.

In this work, 3D printed tumours with diverse shapes, dimensions and locations were inserted into the lungs. The foam compression reproduced a realistic lung motion with an offset depending on the z-position. Without any breath holding or correction, the scotch-yoke mechanism replicated well patient breathing motion and resulted in motion artefacts seen in patient CT images. Casper can therefore be used to generate realistic data to test motion correction and 4D CT image reconstruction. However, the complex-shaped tumour was located close to the spine in the foam lung area subject to small movement. This could be improved by placing the tumour in a different location. Even with the connection mechanism displacement set to 2 cm, the resulting tumour offset was inferior to 0.7 cm for the tumour with the highest offset. However, increasing the motion range that is tailored to the target application could be easily achieved and selected as an acquisition parameter.

While Casper was designed for CT imaging, with some minor modifications it could be used for magnetic resonance imaging (MRI) as well as positron emission tomography (PET) and single-photon emission computed tomography (SPECT) imaging. In addition, the tumours could hold dosimeters, such as thermoluminescent detectors (TLDs), radiochromic films or scintillators and the dose to the moving lung tumour for various radiation treatment delivery techniques could be assessed.

4.3 Cost breakdown

The cost of creating one phantom can be seen below in **Table 4.**

**Table 4**: Manufacturing costs of the Casper phantom (Canadian dollars).

| Item | Price | Where |
|---|---|---|
| Raw materials | 1195.02 | Industrial Paints and Plastics |
| 3D printing | 700 | University of Victoria engineering department |
| Breathing motor | 115.02 | McMaster |
| Sanding equipment | 110.22 | Other |
| Extra machining cost | 18.82 | University of Victoria Machine Shop |
| TOTAL | 2139.08 | |

If the user is attempting to create more than one phantom, several parts and materials can be reused subsequently reducing the cost of each additional phantom. As for the labour of these phantoms, our laboratory resources were used requiring little to no cost. Reproducing this phantom design manufacturing would require to either have or pay for the labour of a 3D printer, laser cutting table, and soldering set. Additionally, if the improvements previously mentioned in the discussion section were applied, the cost would change accordingly.

5. **Conclusions**

We have developed Casper, an anthropomorphic chest phantom with deformable lungs with tumours that is capable of mimicking human breathing motion. Casper is low-cost and made from materials that have similar x-ray attenuation properties to human tissues and it can become a useful tool for x-ray imaging research. All manufacturing data can be downloaded at http://web.uvic.ca/~bazalova/Casper.

**Acknowledgements**

This work was partially funded by an NSERC Discovery Grant and CGS D scholarship, by the Canada Research Chairs program, Canada Foundation for Innovation and British Columbia Knowledge and Development Fund. We would like to thank Teaghan O'Briain for his help with XCAT phantom processing and to Clay Lindsay for help with CT scanning.

.